\documentclass[12pt,a4paper]{article}

\usepackage[english]{babel}
\usepackage{amsmath,amsthm,amssymb}
\usepackage{bbm,graphicx,psfrag,cite}

\numberwithin{equation}{section}

\newcommand{\Id}{\mathbbm{1}}
\newcommand{\Or}{\mathcal{O}}
\newcommand{\Z}{\mathbbm{Z}}
\newcommand{\N}{\mathbbm{N}}
\newcommand{\R}{\mathbbm{R}}
\newcommand{\Pb}{\mathbbm{P}}
\newcommand{\dx}{\mathrm{d}}
\renewcommand{\Re}{\mathrm{Re}}

\newcommand{\Ai}{\mathrm{Ai}}
\newcommand{\Bt}{{\cal A}_{2\to 1}}
\newcommand{\Ap}{{\cal A}_{\rm 2}}
\newcommand{\Af}{{\cal A}_{\rm 1}}
\newcommand{\e}{\varepsilon}

\newcommand{\Dt}[2]{\frac{\dx #1}{\dx #2}}
\newcommand{\I}{{\rm i}}
\newcommand{\Hilb}{{\cal H}}

\newtheorem{prop}{Proposition}
\newtheorem{thm}[prop]{Theorem}

\newtheorem{defin}[prop]{Definition}

\newenvironment{proofOF}[2]{\removelastskip\vspace{6pt}\noindent
 {\it Proof of #1.}~\rm#2}{\qed \par\vspace{6pt}}

\title{Transition between Airy$_1$ and Airy$_2$ processes and TASEP fluctuations}
\author{Alexei Borodin\thanks{California Institute of Technology, e-mail: borodin@caltech.edu},
Patrik L. Ferrari\thanks{Weierstrass Institute (WIAS), Berlin, e-mail: ferrari@wias-berlin.de},\\
Tomohiro Sasamoto\thanks{Chiba University, e-mail: sasamoto@math.s.chiba-u.ac.jp}}

\date{March 6, 2007}

\begin{document}
\maketitle \sloppy

\begin{abstract}
We consider the totally asymmetric simple exclusion process, a model in the KPZ universality class. We focus on the fluctuations of particle positions starting with certain deterministic initial conditions. For large time $t$, one has regions with constant and linearly decreasing density. The fluctuations on these two regions are given by the Airy$_1$ and Airy$_2$ processes, whose one-point distributions are the GOE and GUE Tracy-Widom distributions of random matrix theory. In this paper we analyze the transition region between these two regimes and obtain the transition process. Its one-point distribution is a new interpolation between GOE and GUE edge distributions.
\end{abstract}

\section{Introduction}
In the search of universal limit distribution functions and limit processes, we consider the KPZ universality class (KPZ for Kardar-Parisi-Zhang) originally introduced for stochastic growth models~\cite{KPZ86}. For growth in $1+1$ dimensions the scaling exponents of fluctuations, $1/3$, and correlations, $2/3$, can be (non rigorously) determined by some involved arguments, see e.g.\ \cite{KS92} for an extended discussion. However, to get more insights into the limit laws and limit processes, one is led to consider solvable models in the universality class.

One such model is the polynuclear growth (PNG) model in discrete time,
which has two interesting limits, where new processes have been
discovered. The first limit is the continuous time PNG model, for which
it has been shown that the surface growing in a droplet shape is,
in the large time limit, governed by the Airy$_2$
process~\cite{PS02}. The second one is the totally asymmetric simple
exclusion process (TASEP), in which quite recently the limit process of
the particles positions starting from a periodic initial conditions
has been unravelled and called Airy$_1$ process~\cite{Sas05,BFPS06}.
In the surface growth picture this corresponds to the flat initial
conditions.

\begin{figure}[t!]
\begin{center}
\psfrag{x}[]{$x$}
\psfrag{t}[]{$t$}
\psfrag{t4}[]{$\tfrac{t}{4}$}
\psfrag{12}[]{$\tfrac12$}
\psfrag{rho}[l]{$\rho(x)$}
\psfrag{k}[l]{$k$}
\psfrag{xk+k}[]{$x_k(t)+k$}
\psfrag{A1}[l]{$\Af$}
\psfrag{A12}[l]{$\Bt$}
\psfrag{A2}[l]{$\Ap$}
\psfrag{Min}[c]{$\textrm{GUE}(n)$}
\includegraphics[height=4.5cm]{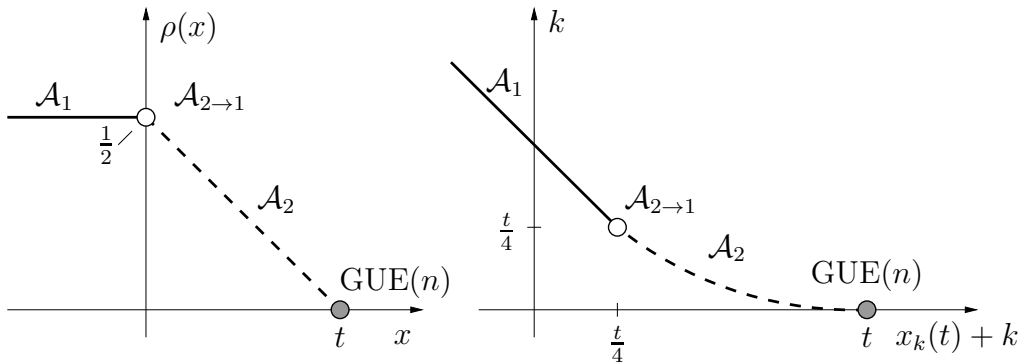}
\caption{Left: the density $\rho$ for large time $t$ is linearly decreasing from $(0,1/2)$ to $(t,0)$. Right: The limit shape in an associated growth, obtained from the density ($k\in\N$ is the label of the particle which starts at $-2k$ and $x_k(t)$ is its position at time $t$).}\label{Fig4Regimes}
\end{center}
\end{figure}

If $x_k(t)$ denotes the position of particle with label $k$, one of the usual geometric representation of the TASEP in terms of surface growth is obtained by the graph $\{(k,x_k(t)+k)\}$, see Figure~\ref{Fig4Regimes} (right) and also e.g.\ \cite{Fer07}. By universality it is expected that the limit process in one-dimensional KPZ growth is the Airy$_2$ process for the curved regions of the limit shape, and the Airy$_1$ process for the flat parts. However initial conditions can easily generate limit shapes which have both curved and flat regions. Therefore there exist transition regions where the limit shape smoothly changes between curved and flat.

The novelty of this paper is the analysis of this transition region in the framework of the TASEP. The observables we consider are positions of several particles at time $t$. In~\cite{Jo03b}, step-initial conditions (particles starting from $\Z_-$) have been considered from the perspective of a growth model and it was proved that the Airy$_2$ process appears in the large time limit. In~\cite{BFPS06,BFP06} we considered periodic initial conditions (particles starting from $d\Z$, $d=2,3,\ldots$) and obtained the Airy$_1$ process as the limit process.

To obtain both regimes and the transition region, we consider in this paper particles starting from $2\Z_-$ as in~\cite{Sas05}. There are four regions of interest as illustrated in Figure~\ref{Fig4Regimes} (left).\\[0.5em]
\textbf{(1) Constant density region.} The limit process of particle positions is given by the Airy$_1$ process $\Af$. In particular, the one-point distribution is $F_1(2^{2/3}s)$, with $F_1$ being the GOE Tracy-Widom distribution.\\[0.5em]
\textbf{(2) Linearly decreasing density region.} The limit process is the Airy$_2$ process $\Ap$, which has $F_2(s)$ as one-point distribution, with $F_2$ being the GUE Tracy-Widom distribution.\\[0.5em]
\textbf{(3) Finite distance from the right-most particle.} There the particle positions are described via the GUE-minor kernel~\cite{JN06}. In particular, the $n$-th right-most particle is distributed as the largest eigenvalue of the $n$-particle GUE ensemble.\\[0.5em]
\textbf{(4) The transition region between (1) and (2).} The fluctuations are governed by a new process obtained in this paper: the transition process Airy$_{2\to 1}$, which we denote $\Bt$. In particular, the one-point distribution interpolates between $F_2(s)$ and $F_1(2^{2/3}s)$ and the transition region has width which scales in time as $t^{2/3}$.

The analysis is done by using the framework of \emph{signed}
determinantal point processes introduced in~\cite{BFPS06}.
This new approach allows us to analyze all four regions for our
initial conditions. This is contrasted to the previously used determinantal
point process issued by the RSK construction, by which only the step initial
condition or its variants could be analyzed~\cite{Jo03b,SI03,SI04,SI07,RB04,FS05a,PS02,BO04}.
We explain how the analysis has to be done for all 4 cases, but the complete asymptotic analysis
is presented only for the transition region, the technically most
difficult one, and the really new result of this paper. The result
is a process, $\Bt$, interpolating between the Airy$_2$ and the
Airy$_1$ processes. For more details about the Airy processes,
see the review~\cite{Fer07}.

The transition we discovered is not the first one between some GUE and GOE type distributions, but it seems to be different from the one previously known for random matrices, non-colliding Brownian motions with open boundary condition and so on~\cite{FNH99,KNT03,NKT02,SI03}. The main differences are the following. On the natural scale of the problems considered, the final distribution is $F_1(2^{2/3}s)$ for our case and $F_1(s)$ in the previous case. Secondly, in the previous case, the GOE-type distribution appears at a single point, while in our case, the GOE-type distribution is on an extended region. Moreover, our transition smoothly interpolates between $F_2(s)$ and $F_1(2^{2/3}s)$, which is not the case for the other transition. In principle, we can not however yet exclude that by a change of variable, with both the rescaling of fluctuations and spatial correlations, the two transitions map one to the other.

\emph{Outline.} In Section~\ref{SectModel} we define the model we analyze and state the results. In Section~\ref{SectKernel} we explain the finite time result and set the scaling limit. In Section~\ref{SectAnalysis} we do the complete asymptotic analysis for the transition region and in Section~\ref{SectThreeRegions} we explain how to do the analysis for the other cases. Finally, we present an explicit form of the transition kernel in terms of Airy functions in Appendix~\ref{AppExplicitKernel} and we explain the correctness of the Fredholm determinants involved in~\ref{AppTraceClass}.

\subsubsection*{Acknowledgments} A. Borodin was partially supported by the NSF grant DMS-0402047
and the CRDF grant RIM1-2622-ST-04. P.L. Ferrari would like to thank for the support of the German-Japanese cooperation project of German National Foundation, DFG 446JAP113/325/0-1. The work of T. Sasamoto is supported by the Grant-in-Aid for Young Scientists (B), the Ministry of Education, Culture, Sports, Science and Technology, Japan.

\section{Model and results}\label{SectModel}
In this paper we consider the continuous-time totally asymmetric simple exclusion process (TASEP) on $\Z$. At any given time $t$, every site $j\in\Z$ can be occupied at most by one particle. Thus a configuration of the TASEP can be described by $\eta=\{\eta_j,j\in\Z|\eta_j\in\{0,1\}\}\in\Omega=\{0,1\}^\Z$. $\eta_j$ is called the \emph{occupation variable} of site $j$, which is defined by $\eta_j=1$ if site $j$ is occupied and $\eta_j=0$ if site $j$ is empty.

The dynamics of the TASEP is defined as follows. Particles jump on the neighboring right site with rate $1$ provided that the site is empty. This means that jumps are independent of each other and are performed after an exponential waiting time with mean $1$. More precisely, let $f$: $\Omega\to \R$ be a function depending only on a finite number of $\eta_j$'s. Then the backward generator of the TASEP is given by
\begin{equation}\label{1.1}
Lf(\eta)=\sum_{j\in\Z}\eta_j(1-\eta_{j+1})\big(f(\eta^{j,j+1})-f(\eta)\big).
\end{equation}
Here $\eta^{j,j+1}$ denotes the configuration $\eta$ with the occupations at sites $j$ and \mbox{$j+1$} interchanged. The semigroup $e^{Lt}$ is well-defined as acting on bounded and continuous functions on $\Omega$. $e^{Lt}$ is the transition probability of the TASEP~\cite{Li99}.

We denote by $x_k(t)$ the position of the particle number $k$ at time $t$. As initial condition we consider particles starting from $2\Z_-$, i.e., $x_k(0)=-2k$ for $k=1,2,\ldots$.
On the macroscopic level, the limit particle density $u(\xi)$ is given by
\begin{equation}\label{eqMacroDensity}
u(\xi)=\frac{\mathrm{d}}{\mathrm{d}\xi}\lim_{t\to\infty} \frac1t\mathbb{E}\big(\#(k: x_k(t)\geq \xi t)\big)=
\left\{\begin{array}{ll}
1/2,& \xi<0,\\
1/2-\xi/2,&\xi\in [0,1],\\
0,&\xi>1.
\end{array}\right.
\end{equation}
Thus for large time $t$ the expected number of particle at site $x$, $\rho(x)$, is close to $u(x/t)$, see Figure~\ref{Fig4Regimes}.

As observables we consider the positions of finite subsets of particles, $\{x_i(t),i\in I\}$ for some $I\subset \N$, $|I|<\infty$. The scaling limits we have to take depend on which of the four regions described in the Introduction we focus on, see also Figure~\ref{Fig4Regimes}. The main result of this paper is the description of the large time fluctuations in the transition region, which now we describe.

\subsubsection*{(4) Transition region: the $\Bt$ process}
The transition region has width of order $t^{2/3}$, which is indicated by the fact that the index of the particles which at time $t$ are around $x=0$ fluctuates on the $t^{2/3}$ scale around the macroscopic value $t/4$. Therefore we set
\begin{equation}
n(\tau,t)=[t/4+\tau (t/2)^{2/3}].
\end{equation}
The density (\ref{eqMacroDensity}) changes in the transition region. The limit density can be used to determine, on the macroscopic scale, the expected location at time $t$ of a particle with index $n(a)=[t/4+a t]$ (of course $a\geq-1/4$). The result is then
\begin{equation}\label{eqMacroPosition}
\lim_{t\to\infty} \frac{x_{n(a)}}{t}=
\left\{\begin{array}{ll}
1-\sqrt{1+4 a},& a \in [-1/4,0],\\
-2a,& a\geq 0.
\end{array}\right.
\end{equation}
By using (\ref{eqMacroPosition}) with $a t = \tau (t/2)^{2/3}$ we are led to define the rescaled process of particle positions by
\begin{equation}\label{eqRescaling}
\tau \mapsto X_t(\tau)=\frac{x_{n(\tau,t)}(t)-(-2\tau(t/2)^{2/3}+\min\{0,\tau\}^2 (t/2)^{1/3})}{-(t/2)^{1/3}}.
\end{equation}

The main result of this paper is the convergence of $X_t(\tau)$ to the transition process $\Bt$ defined below.
\begin{defin}[The Airy$_{2\to 1}$ process]\label{Defin}
Let us set
\begin{equation}
\tilde s_i=\left\{\begin{array}{ll} s_i,& \tau_i\geq 0,\\ s_i-\tau_i^2,&  \tau_i\leq 0. \end{array}\right.
\end{equation}
and define the transition kernel
\begin{eqnarray}\label{eqKCompleteInfinity}
& &K_\infty(\tau_1,s_1;\tau_2,s_2)=-\frac{1}{\sqrt{4\pi(\tau_2-\tau_1)}}\exp\left(-\frac{(\tilde s_2-\tilde s_1)^2}{4(\tau_2-\tau_1)}\right)\Id(\tau_2>\tau_1)\nonumber \\
& &+\frac{1}{(2\pi \I)^2} \int_{\gamma_+}\dx w \int_{\gamma_-}\dx z
\frac{e^{w^3/3+\tau_2 w^2-\tilde s_2 w}}{e^{z^3/3+\tau_1 z^2-\tilde s_1 z}} \frac{2w}{(z-w)(z+w)}
\end{eqnarray}
with the paths $\gamma_+,\gamma_-$ satisfying $-\gamma_+\subset\gamma_-$ with $\gamma_+:e^{\I\phi_+}\infty\to e^{-\I\phi_+}\infty$, \mbox{$\gamma_-:e^{-\I\phi_-}\infty\to e^{\I\phi_-}\infty$} for some $\phi_+\in (\pi/3,\pi/2)$, $\phi_-\in (\pi/2,\pi-\phi_+)$, see Figure~\ref{FigPathsDefin}.
\begin{figure}[t!]
\begin{center}
\psfrag{pi3}[l]{$\pi/3$}
\psfrag{g+}[l]{$\gamma_+$}
\psfrag{g-}[r]{$\gamma_-$}
\psfrag{-g+}[r]{$-\gamma_+$}
\includegraphics[height=5cm]{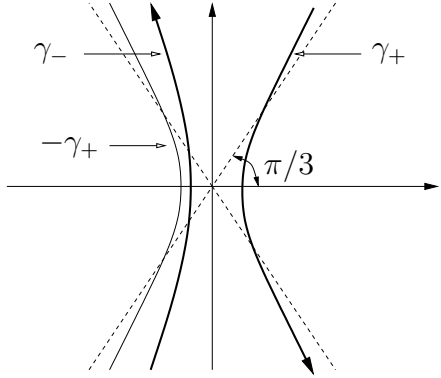}
\caption{An illustration of the paths $\gamma_+$ and $\gamma_-$ in Definition~\ref{Defin}.}\label{FigPathsDefin}
\end{center}
\end{figure}

The \emph{Airy$_{2\to 1}$ process}, $\Bt$, is the process with $m$-point joint distributions at $\tau_1<\tau_2<\ldots<\tau_m$ given by the Fredholm determinant
\begin{equation}\label{eqTransAiryProcess}
\Pb\bigg(\bigcap_{k=1}^m\{\Bt(\tau_k)\leq s_k\}\bigg)=\det(\Id-\chi_s K_\infty \chi_s)_{L^2(\{\tau_1,\ldots,\tau_m\}\times \R)}
\end{equation}
where $\chi_s(\tau_k,x)=\Id(x>s_k)$. An explicit expression for $K_\infty$ in terms of Airy functions can be found in Appendix~\ref{AppExplicitKernel}.
\end{defin}

\noindent\textbf{Remarks:} $\Bt(t+\tau)$ becomes $2^{-1/3}\Af(2^{2/3}\tau)$ as $t\to\infty$ and $\Ap(\tau)$ when $t\to-\infty$. The Fredholm determinant in (\ref{eqTransAiryProcess}) is well defined because, as proven in Proposition~\ref{PropTraceClass} of Appendix~\ref{AppTraceClass}, there exists a conjugate kernel of $\chi_s K_\infty \chi_s$ which is trace-class on $\Hilb=L^2(\{\tau_1,\ldots,\tau_m\}\times \R)$.

Now we can state precisely our main Theorem.
\begin{thm}\label{ThmMain}
The convergence of $X_t$ to the transition process $\Bt$,
\begin{equation}
\lim_{t\to\infty} X_t(\tau) = \Bt(\tau),
\end{equation}
holds in the sense of finite-dimensional distributions.
\end{thm}

\noindent\textbf{A remark on initial conditions.} In this work as well as in many of the previous papers in the field, the situations analyzed with deterministic initial conditions might look quite peculiar: step-initial conditions~\cite{Jo03}, periodic with period $2$ or more~\cite{BFPS06, BFP06}. However, it is intuitively clear that small perturbations of the initial conditions do not affect the large time behavior. This is indeed the case by a coupling argument.

Consider two TASEP initial conditions of $N$ particles, $X(0)=\{x_N(0)<\ldots<x_2(0)<x_1(0)\}$, $Z(0)=\{z_N(0)<\ldots<z_2(0)<z_1(0)\}$ with $X(0)\leq Z(0)$ meaning $x_k(0)\leq z_k(0)$, $k=1,\ldots,N$. By a standard coupling argument, see e.g.~\cite{Lig76}, for any subset $I\subset \{1,\ldots,N\}$,
\begin{equation}\label{eqCoupling}
\Pb(\{x_i(t)\leq a_i,i\in I\})\geq \Pb(\{z_i(t)\leq a_i,i\in I\}).
\end{equation}

We can apply (\ref{eqCoupling}) to our case to show that the limit result is unchanged if we do any bounded perturbation of the initial condition. In Theorem~\ref{ThmMain} we started with initial conditions $x_i(0)=-2i$. Consider any other initial condition $Z=\{z_i(0)\}$ and define
\begin{equation}
M=\max\{|x_i(0)-z_i(0)|\}.
\end{equation}
Then, by (\ref{eqCoupling}), we have
\begin{eqnarray}\label{eq2.12}
\Pb(\{x_i(t)\leq a_i+M,i\in I\})&\geq& \Pb(\{z_i(t)\leq a_i,i\in I\}) \nonumber \\
&\geq& \Pb(\{x_i(t)\leq a_i-M,i\in I\}).
\end{eqnarray}
In the scaling limit (\ref{eqRescaling}), the first and last term in (\ref{eq2.12}) have the same limit as $t\to\infty$ as long as $\lim_{t\to\infty} M/t^{1/3}=0$. This holds in particular if $Z$ is any bounded perturbation of $X$, i.e., if $M<\infty$ is independent of $t$.

For completeness we state the results in the other three regions. In Section~\ref{SectThreeRegions} we outline how the asymptotic analysis for the transition region has to be modified in order to obtain the results. The scaling is obtained using (\ref{eqMacroPosition}).
\subsubsection*{(1) Fixed particle number: $\textrm{GUE}(n)$ minors}
Consider particles with index not rescaled in time, i.e.,
\begin{equation}
n\text{ of order one},
\end{equation}
and the rescaled random variables
\begin{equation}
X_t(n)=\frac{x_{n}(t)-t}{-\sqrt{2t}}.
\end{equation}
Then, in the $t\to\infty$ limit, one gets the $\textrm{GUE-minors}(n)$ given in~\cite{JN06},
\begin{equation}\label{eqGUEn}
\lim_{t\to\infty} X_t(n) = \textrm{GUE-minors}(n).
\end{equation}

\subsubsection*{(2) Linearly decreasing density region: Airy$_2$ process, $\Ap$ }
For $0<\alpha<1$, define
\begin{equation}
n(\tau,t)=[\alpha t/4+\tau (t/2)^{2/3}],
\end{equation}
and the rescaled process
\begin{equation}
\tau \mapsto X_t(\tau)=\frac{x_{n(\tau,t)}(t)-((1-\sqrt{\alpha})t-2\tau \alpha^{-1/2} (t/2)^{2/3}+\tau^2 \alpha^{-3/2}(t/2)^{1/3})}{-(t/2)^{1/3}}.
\end{equation}
Then in the $t\to\infty$ limit, one gets
\begin{equation}\label{eqCurved}
\lim_{t\to\infty} X_t(\tau) = \frac{(2-\sqrt{\alpha})^{2/3}}{\alpha^{1/6}} \Ap(\tau \alpha^{2/3}(2-\sqrt{\alpha})^{1/3}).
\end{equation}

\subsubsection*{(3) Constant density region: Airy$_1$ process, $\Af$}
For $\alpha>1$,
\begin{equation}
n(\tau,t)=[\alpha t/4+\tau (t/2)^{2/3}],
\end{equation}
and the rescaled process variables
\begin{equation}
\tau \mapsto X_t(\tau)=\frac{x_{n(\tau,t)}(t)-((1-\alpha)t/2-2\tau(t/2)^{2/3})}{-(t/2)^{1/3}}.
\end{equation}
Then in the $t\to\infty$ limit, one gets
\begin{equation}\label{eqFlat}
\lim_{t\to\infty} X_t(\tau) = 2^{1/3} \Af(2^{2/3} \tau).
\end{equation}

\section{Kernel and its scaling limit}\label{SectKernel}
In this section we derive the expression of the joint distributions of particle positions and then set the proper scaling limit.

Consider $N$ particles starting at time $t=0$ at positions $x_k(0)=-2k$, $k=1,\ldots,N$. In Theorem 2.1 of~\cite{BFPS06} we proved that the joint distribution of the positions of the particles are given by a Fredholm determinant. The kernel is determined via a certain orthogonalization, which for our initial conditions has been made in Lemma 4.1 of~\cite{BFPS06} (with $z=x+2n-2N$ replaced by $z=x+2n$). Once the orthogonalization is made, one can compute the kernel which is (4.11) of~\cite{BFPS06} (with $z_i=x_i+2n_i-2N$ replaced by $z_i=x_i+2n_i$). This is summarized in Proposition~\ref{PropKernel}.

\begin{prop}\label{PropKernel}
Let particle with label $i$ start at $x_i(0)=-2i$, $i=1,\ldots,N$. At time $t$, the particles are at positions $x_i$. Let $\sigma(1)<\sigma(2)<\ldots<\sigma(m)$ be the indices of $m$ out of the $N$ particles. The joint distribution of their positions $x_{\sigma(k)}(t)$ is given by
\begin{equation}
\Pb\Big(\bigcap_{k=1}^m \big\{x_{\sigma(k)}(t) \geq a_k\big\}\Big)=
\det(\Id-\chi_a K_t\chi_a)_{\ell^2(\{\sigma(1),\ldots,\sigma(m)\}\times\Z)}
\end{equation}
where $\chi_a(\sigma(k),x)=\Id(x<a_k)$. The kernel $K_t$ is given by
\begin{equation}
K_t(n_1,x_1;n_2,x_2)=-\binom{x_1-x_2-1}{n_2-n_1-1}\Id_{[n_2>n_1]}+\widehat{K}_t(n_1,x_1;n_2,x_2),
\end{equation}
where
\begin{eqnarray}
\widehat{K}_t(n_1,x_1;n_2,x_2)&=&\frac{(-1)^{n_1-n_2}}{(2\pi \I)^2}\oint_{\Gamma_0}\dx v \oint_{\Gamma_{-1}} \dx u \frac{e^{-vt}(1+v)^{x_2+n_2}}{v^{n_2}}\\
& & \frac{e^{ut}u^{n_1}}{(1+u)^{x_1+n_1+1}}\frac{1+2v}{(u-v)(1+u+v)}\nonumber
\end{eqnarray}
where $\Gamma_0$, resp.\ $\Gamma_{-1}$, is any simple loop, anticlockwise oriented, which includes the pole at $v=0$, resp.\ $u=-1$, satisfying $-1-\Gamma_0\subset \Gamma_{-1}$, i.e., all the points of $-1-\Gamma_0$ lie inside the loop $\Gamma_{-1}$.
\end{prop}

In order to prove Theorem~\ref{ThmMain}, we need to focus at particles with number $n_i$ close to $t/4$ since these particles will be in the transition region at time $t$. The transition region has width which scales as $t^{2/3}$. The limit density is constant to the left of the transition region and it is decreasing linearly to the right of it. Therefore, the scaling limit used to prove the main theorem is
\begin{eqnarray}\label{eqscaling}
n_i&=&[t/4+\tau_i (t/2)^{2/3}],\nonumber \\
x_i&=&[-2\tau_i(t/2)^{2/3}-\tilde s_i (t/2)^{1/3}],
\end{eqnarray}
where
\begin{equation}
\tilde s_i=\left\{\begin{array}{ll} s_i,& \tau_i\geq 0,\\ s_i-\tau_i^2,&  \tau_i\leq 0. \end{array}\right.
\end{equation}
As a consequence the rescaled kernel writes
\begin{equation}
K_t^{\rm resc}(\tau_1,s_1;\tau_2,s_2)=K_t(n_1,x_1;n_2,x_2) (t/2)^{1/3} 2^{x_2-x_1}
\end{equation}
where $2^{x_2-x_1}$ is just a conjugation so that the kernel has a proper limit. We denote by
$\widehat{K}_t^{\rm resc}$ the term of the rescaled kernel without the binomial contribution, which then writes
\begin{eqnarray}\label{eq6}
& &\widehat{K}_t^{\rm resc}(\tau_1,s_1;\tau_2,s_2)=(t/2)^{1/3}\frac{1}{(2\pi\I)^2}\oint_{\Gamma_0}\dx v \oint_{\Gamma_{-1}}\dx u \frac{1+2v}{(u-v)(1+u+v)} \nonumber\\
& &\times \frac{\exp(t f_0(v)+(t/2)^{2/3}\tau_2 f_1(v)+(t/2)^{1/3}\tilde s_2 f_2(v))}
{\exp(t f_0(u)+(t/2)^{2/3}\tau_1 f_1(u)+(t/2)^{1/3}\tilde s_1 f_2(u)+f_3(u))},
\end{eqnarray}
where the functions $f_i$ are given by
\begin{eqnarray}\label{eq7}
f_0(v)&=&-v+\tfrac14 \ln((1+v)/v),\nonumber \\
f_1(v)&=&-\ln(-4v(1+v)),\nonumber \\
f_2(v)&=&-\ln(2(1+v)),\nonumber \\
f_3(v)&=&\ln(1+v).
\end{eqnarray}
From now on the $\tau_i$'s are some \emph{fixed} values. With this preparation we can proceed to the asymptotic analysis needed to prove Theorem~\ref{ThmMain}.

\section{Asymptotic analysis}\label{SectAnalysis}

\begin{proofOF}{Theorem~\ref{ThmMain}}
The proof of Theorem~\ref{ThmMain} is identical to the one of Theorem 2.5 in~\cite{BFP06}, provided the following Propositions~\ref{PropUniformConvergence}, \ref{PropModerateDeviations}, \ref{PropLargeDeviations}, \ref{PropUniformConvergenceBinomial}, and \ref{PropBoundBinomial} (convergence on bounded sets and large deviations bounds) hold.
\end{proofOF}

\begin{prop}[Uniform convergence on bounded sets]\label{PropUniformConvergence}
Fix any $L>0$ and $x_i,s_i$ with the scaling (\ref{eqscaling}). Then, uniformly for $(s_1,s_2)\in [-L,L]^2$,
\begin{equation}
\lim_{t\to\infty} \widehat{K}_t^{\rm resc}(n_1,x_1;n_2,x_2)=\widehat K_{\infty}^{\rm resc}(\tau_1,s_1;\tau_2,s_2)
\end{equation}
where
\begin{equation}\label{eqKinfinity}
\widehat{K}_\infty(\tau_1,s_1;\tau_2,s_2)=\frac{1}{(2\pi \I)^2} \int_{\gamma_+}\dx w \int_{\gamma_-}\dx z
\frac{e^{w^3/3+\tau_2 w^2-\tilde s_2 w}}{e^{z^3/3+\tau_1 z^2-\tilde s_1 z}} \frac{2w}{(z-w)(z+w)}
\end{equation}
with the paths $\gamma_+,\gamma_-$ satisfying $-\gamma_+\subset\gamma_-$ with $\gamma_+:e^{\I\phi_+}\infty\to e^{-\I\phi_+}\infty$, $\gamma_-:e^{-\I\phi_-}\infty\to e^{\I\phi_-}\infty$ for some $\phi_+\in (\pi/3,\pi/2)$, $\phi_-\in (\pi/2,\pi-\phi_+)$.
\end{prop}
\begin{proof}
The first step is to control the contribution away from the critical point given by
\begin{equation}
\Dt{f_0(v)}{v}=-\frac{(1+2v)^2}{4v(1+v)}=0\quad \iff \quad v=-1/2.
\end{equation}
If we write $v=x+\I y$, $x,y\in\R$, then we can analyze
\begin{equation}
\Re(f_0(v)-f_0(-1/2))=-(x+1/2)+\tfrac18\ln(((1+x)^2+y^2)/(x^2+y^2)).
\end{equation}
This expression equals zero for\\
a) $x=-1/2$, $y\in\R$,\\
b) $y=\pm g(x)$, with $g(x)=\sqrt{\frac{1+2x+x^2(1-e^{8x+4})}{e^{8x+4}-1}}$.\\
If is easy to see that the solutions $\pm g(x)$ are symmetric with respect to \mbox{$v=-1/2$} and they go around $-1$ and $0$ once. Moreover, the loops leave the critical point $v=-1/2$ in the directions $e^{\pm \I\pi/6}$ and $e^{\pm \I 5\pi/6}$, see Figure~\ref{FigZeros}.
\begin{figure}[h!]
\begin{center}
\psfrag{-1}[c]{$-1$}
\psfrag{-0.5}[c]{$-0.5$}
\psfrag{ 0}[c]{$0$}
\psfrag{-0.2}[c]{$-0.2$}
\psfrag{ 0.2}[c]{$0.2$}
\psfrag{D1}[c]{$D_1$}
\psfrag{D2}[c]{$D_2$}
\psfrag{D3}[c]{$D_3$}
\psfrag{D4}[c]{$D_4$}
\psfrag{x}[c]{$x$}
\psfrag{y}[c]{$y$}
\includegraphics[height=6cm]{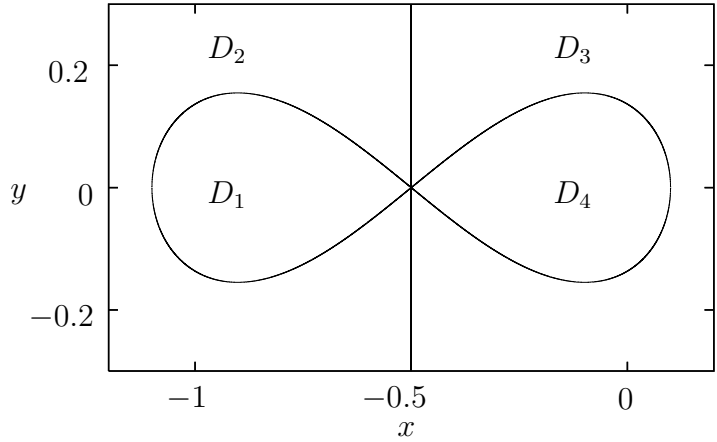}
\caption{The signum of $\Re(f_0(x+\I y)-f_0(-1/2))$ is positive in $D_2$ and $D_4$ and negative in $D_1$ and $D_3$.}
\label{FigZeros}
\end{center}
\end{figure}
We denote by $D_1,\ldots,D_4$ the following regions: $D_1$ is the region enclosed by $\pm g$ around $-1$, $D_2$ is the rest with real part less than $-1/2$, $D_4$ is the symmetric image w.r.t. $-1/2$ of $D_1$ and $D_3$ of $D_2$, see Figure~\ref{FigZeros}. Then $\Gamma_0$ can be chosen to be any simple anticlockwise oriented finite length path staying in $D_3$ and, similarly, $\Gamma_{-1}$ is chosen to stay in $D_2$ (except at $v=-1/2$). The constraint $-1-\Gamma_0\subset \Gamma_{-1}$ is easily satisfied except that for $\Gamma_0$ we have to go through $D_4$ too, very close to $v=-1/2$. Moreover, we can take $\Gamma_0$ leaving from $-1/2$ with an angle between $-\pi/6$ and $-\pi/3$. Similarly, $\Gamma_{-1}$ leaves in the direction from $2\pi/3$ and $5\pi/6$. This will simplify the argument for moderate and large deviations.
\begin{figure}[h!]
\begin{center}
\psfrag{G1}[l]{$\Gamma_{-1}$}
\psfrag{G11}[l]{$-1-\Gamma_{-1}$}
\psfrag{G0}[l]{$\Gamma_{0}$}
\psfrag{pi}[l]{$\pi/6$}
\psfrag{delta}[c]{$\delta t^{-1/3}$}
\includegraphics[height=6cm]{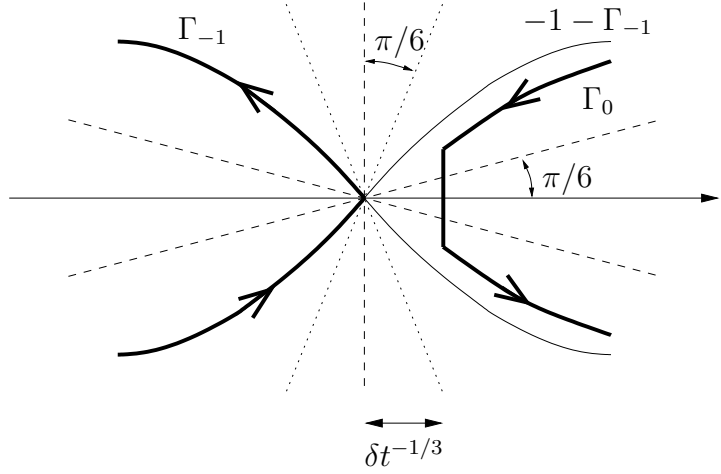}
\caption{The paths $\Gamma_0$ and $\Gamma_{-1}$ close to the critical point $-1/2$. The dashed lines are the zeros of $\Re(f_0(x+\I y)-f_0(-1/2))$.}
\label{FigSteepPaths}
\end{center}
\end{figure}

Let us set $\Gamma_0^\delta=\{v\in\Gamma_0, |v+1/2|\leq \delta\}$ and
 $\Gamma_{-1}^\delta=\{u\in\Gamma_{-1}, |u+1/2|\leq \delta\}$. Then the
 integral is over $\Gamma_0\cup \Gamma_{-1}=\Gamma_0^\delta\cup
 \Gamma_{-1}^\delta+\Sigma$, where $\Sigma$ is the rest of the
 contours. The first step is to bound the integral over $\Sigma$. For
 $0<\delta\ll 1$, we can choose $\Gamma_0$ and $\Gamma_{-1}$ such that,
 for $(u,v)\in \Sigma$, $|u-v|/\delta$ and $|1+u+v|/\delta$ are bounded away from $0$. Then, on $\Sigma$, $\left|\frac{1+2v}{(u-v)(1+u+v)}\right|\leq |u-v|^{-1}+|1+u+v|^{-1}= \Or(1/\delta)$ and, for some $c_0=c_0(\delta)>0$,
\mbox{$\Re(f_0(v)-f_0(-1/2))\leq -c_0$}
and/or $-\Re(f_0(u)-f_0(-1/2))\leq -c_0$.
Thus, the integral over $\Sigma$ can be bounded as
\begin{equation}
c_1 \delta^{-1} t^{1/3} \exp(-c_0 t+\Or(t^{2/3}))
\end{equation}
for some $c_1>0$. For $t$ large enough, both $e^{\Or(t^{2/3})}$ and $c_1 t^{1/3}$ are bounded by $e^{-c_0 t/4}$. Thus, for $t$ large enough, we have the bound
\begin{equation}
\left|\iint_{\Sigma} \cdots\right|\leq \delta^{-1} e^{-c_0 t/2}.
\end{equation}

The second step is to control the integral over $\Gamma_0^\delta\cup\Gamma_{-1}^\delta$. Since $\delta$ is small, we can apply Taylor series expansion on the functions $f_i$ defined in (\ref{eq7}). For this we change variables by setting
\begin{equation}
u=-1/2+U,\quad v=-1/2+V
\end{equation}
and we denote $\gamma_+^\delta=\Gamma_0^\delta+1/2$, $\gamma_-^\delta=\Gamma_{-1}^\delta+1/2$. We have
\begin{eqnarray}
f_0&=&\tfrac12+\I \tfrac{\pi}{4} +\tfrac43 V^3+\Or(V^4),\nonumber \\
f_1&=&4V^2+\Or(V^4),\nonumber \\
f_2&=&-2V+\Or(V^2),\nonumber \\
f_3&=&-\ln(2)+\Or(V).
\end{eqnarray}
Therefore the integral over $\Gamma_0^\delta\cup\Gamma_{-1}^\delta$ is given by
\begin{eqnarray}\label{eq43}
& &\frac{(t/2)^{1/3}}{(2\pi\I)^2}\int_{\gamma_+^\delta}\dx V \int_{\gamma_-^\delta}\dx U \frac{4V}{(U-V)(U+V)} \frac{e^{\frac43 t V^3+(t/2)^{2/3}\tau_2 4V^2-\tilde s_2(t/2)^{1/3} 2V}}{e^{\frac43 t U^3+(t/2)^{2/3}\tau_1 4U^2-\tilde s_1(t/2)^{1/3} 2U}} \nonumber\\
&\times& e^{\Or(t V^4,t^{2/3} V^4,L t^{1/3} V^2,t U^4,t^{2/3} U^4,L t^{1/3} U^2,U)}\nonumber \\
&=&\frac{(t/2)^{1/3}}{(2\pi\I)^2}\int_{\gamma_+^\delta}\dx V \int_{\gamma_-^\delta}\dx U \frac{4V}{(U-V)(U+V)} \frac{e^{\frac43 t V^3+(t/2)^{2/3}\tau_2 4V^2-\tilde s_2(t/2)^{1/3} 2V}}{e^{\frac43 t U^3+(t/2)^{2/3}\tau_1 4U^2-\tilde s_1(t/2)^{1/3} 2U}} \nonumber\\
&&  + \textrm{R}.
\end{eqnarray}
To bound the remainder, $R$, we use $|e^x-1|\leq |x|e^{|x|}$ applied to $x=\Or(\cdots)$. Moreover, note that $\Or(t^{2/3}V^4)$ is dominated by $\Or(tV^4)$. Therefore,
\begin{eqnarray}
|R|&\leq& c_2 t^{1/3} \int_{\gamma_+^\delta}\dx V \int_{\gamma_-^\delta}\dx U \bigg|\frac{4V}{(U-V)(U+V)} \frac{e^{\frac43 t V^3+(t/2)^{2/3}\tau_2 4V^2-\tilde s_2(t/2)^{1/3} 2V}}{e^{\frac43 t U^3+(t/2)^{2/3}\tau_1 4U^2-\tilde s_1(t/2)^{1/3} 2U}} \nonumber \\
&\times & e^{\Or(t V^4,L t^{1/3} V^2,t U^4,L t^{1/3} U^2,U)}\Or(t V^4,L t^{1/3} V^2,t U^4,L t^{1/3} U^2,U)\bigg|.
\end{eqnarray}
At this point we do the change of variables $V=w (4t)^{-1/3}$ and $U=z (4t)^{-1/3}$ and obtain
\begin{eqnarray}
|R|&\leq & c_3 t^{-1/3} \int_{(4t)^{1/3}\gamma_+^\delta}\dx w \int_{(4t)^{1/3} \gamma_-^\delta}\dx z \bigg|\frac{w}{(z-w)(z+w)} \frac{e^{w^3/3+\tau_2 w^2-\tilde s_2 w}}{e^{z^3/3+\tau_1 z^2-\tilde s_1 z}}\nonumber \\
&\times & e^{t^{-1/3}\Or(w^4,Lw^2,z^4,Lz^2,z)} \Or(w^4,Lw^2,z^4,Lz^2,z) \bigg|.
\end{eqnarray}
By choosing $\delta$ small enough, we may assume that $\Or(w^4 t^{-1/3})\ll w^3$, $\Or(z t^{-1/3})\ll 1$, and for $t$ large enough $\Or(L t^{-1/3})\ll 1$. Therefore, the exponential in the integral in the $w$ variable can be bounded by \mbox{$|\exp(\chi_0 w^3/3+\tau_2 \chi_1 w^2-\tilde s_2\chi_2 w)|$} for some $\chi_0,\chi_1,\chi_2$. By choosing $\delta$ small enough, the $\chi$'s can be made as close to $1$ as desired. More importantly, for $\delta$ small, one has $\chi_0>0$. Similar for the variable $z$ for some $\tilde \chi_k$. We have
\begin{eqnarray}\label{eq20}
|R|&\leq & c_3 t^{-1/3} \int_{(4t)^{1/3}\gamma_+^\delta}\dx w \int_{(4t)^{1/3} \gamma_-^\delta}\dx z \bigg|\frac{w}{(z-w)(z+w)} \frac{e^{\chi_0 w^3/3+\tau_2 \chi_1 w^2-\tilde s_2 \chi_2 w}}{e^{\tilde\chi_0 z^3/3+\tau_1\tilde\chi_1 z^2-\tilde s_1 \tilde\chi_2 z}}\nonumber \\
&\times & \Or(w^4,Lw^2,z^4,Lz^2,z) \bigg|.
\end{eqnarray}

The integral in (\ref{eq20}), without the prefactor $t^{-1/3}$, is uniformly bounded in $t$. In fact, the only dependence on $t$ is at the boundaries of the integrals, which are at $\delta e^{\pm\I \theta_+}$ and $\delta e^{\pm \I \theta_-}$ with \mbox{$\theta_+\in (\pi/6,\pi/3)$} and $\theta_-\in (2\pi/3,5\pi/6)$.
The convergence is ensured by the fact that $\Re(w^3)=\delta^3 t \cos(3\theta_+)$, with $\cos(3\theta_+)<0$, and $\Re(-z^3)= -\delta^3 t \cos(3\theta_-)$, with $\cos(3\theta_-)>0$. Thus, the $w^3$ and $z^3$ terms dominate the others at the boundary of the integrals and this domination becomes stronger while $t$ increases. The final result is that, we can set $\delta>0$ small enough and then for $t$ large enough we have
\begin{equation}
|R|\leq c_4 t^{-1/3}.
\end{equation}

The last step is to analyze the first term in r.h.s.\ of (\ref{eq43}). One does the same change of variable as above and gets
\begin{equation}
\frac{1}{(2\pi\I)^2} \int_{(4t)^{1/3}\gamma_+^\delta}\dx w \int_{(4t)^{1/3} \gamma_-^\delta}\dx z \frac{2w}{(z-w)(z+w)} \frac{e^{w^3/3+\tau_2 w^2-\tilde s_2 w}}{e^{z^3/3+\tau_1 z^2-\tilde s_1 z}}.
\end{equation}
We can extend the paths to $t=\infty$ and by doing so we gain the error term of order $\Or(e^{-c_5 \delta^3 t})$ for some $c_5>0$. With this extension the paths satisfy the conditions of $\gamma_+$ and $\gamma_-$ of the Proposition.

Just to summarize, the error term we have accumulated during the above procedure is
\begin{equation}\label{eq22}
\Or(\delta^{-1}e^{-c_0 t/2},c_4 t^{-1/3},e^{-c_5 \delta^3 t}).
\end{equation}
\end{proof}

\begin{prop}[Moderate deviations]\label{PropModerateDeviations}
For any $L$ large enough, $\exists \, \e_0(L)>0$ and $t_0(L)>0$ such that, $\forall \, 0<\e\leq \e_0$ and $t\geq t_0$, the estimate
\begin{equation}
\left|\widehat{K}_t^{\rm resc}(\tau_1,s_1;\tau_2,s_2)\right|\leq e^{-(s_1+s_2)}
\end{equation}
holds for $(s_1,s_2)\in [-L,\e t^{2/3}]^2\setminus [-L,L]^2$.
\end{prop}
\begin{proof}
In this proof we introduce the notation, $\sigma_i=\tilde s_i t^{-2/3} 2^{-1/3} \in (0,\e]$, $i=1,2$. We divide the analysis in the cases $\tilde s_1\geq \tilde s_2$ and $\tilde s_1\leq \tilde s_2$. The strategy is the following. First, for the case $\tilde s_1\geq \tilde s_2$, we choose the same paths $\Gamma_0$ and $\Gamma_{-1}$ as in Proposition~\ref{PropUniformConvergence} except for a small modification close to $v=u=-1/2$. We then see that in the unmodified part of the paths one has the same integral as for the case $\sigma_1=\sigma_2=0$ times a factor which can be simply bounded and gives the needed decay. Then we consider the modified parts of the integration paths and see that the integral over these has also the required decay. Secondly, for the case $\tilde s_1\leq \tilde s_2$, we first modify the condition of the integral since, otherwise, the optimal paths for the exponential can not be followed close to the critical points. The modification produces an extra ter
 m, a residue, which is a simple integral and it can be bounded in a similar way.

\textbf{Case $\sigma_1\leq \sigma_2$.} The paths $\Gamma_0$ and $\Gamma_{-1}$ as represented in Figure~\ref{FigPathsModerate}.
\begin{figure}[h!]
\begin{center}
\psfrag{Gv}[l]{$\Gamma_{\rm vert}$}
\psfrag{G1}[c]{$\Gamma_{-1}$}
\psfrag{G0}[c]{$\Gamma_{0}$}
\psfrag{-1}[c]{$-1$}
\psfrag{0}[l]{$0$}
\psfrag{a}[r]{$\frac{\sqrt{\sigma_1}}{4}$}
\psfrag{b}[l]{$\frac{\sqrt{\sigma_2}}{2}$}
\includegraphics[height=6cm]{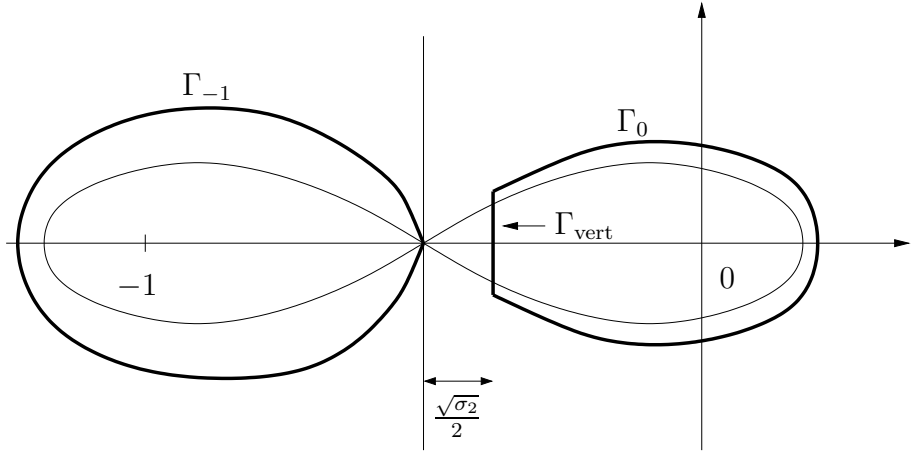}
\caption{The paths $\Gamma_0$ and $\Gamma_{-1}$ used to obtain the bound in the moderate deviations regime for $\sigma_1\leq\sigma_2$.}
\label{FigPathsModerate}
\end{center}
\end{figure}

The modification with respect to the ones in Proposition~\ref{PropUniformConvergence} is just one vertical piece, given by $\Gamma_{\rm vert}=\{-1/2+\sqrt{\sigma_2}(1+\I \xi)/2, \xi\in[-a,a]\}$ for some $a\in (1/\sqrt{3},\sqrt{3})$.

With respect to the case $\sigma_1=\sigma_2=0$, the integrand in the integral representation of the kernel $\widehat{K}_t^{\rm resc}$, see (\ref{eq6}), has the extra factor
\begin{equation}\label{eq23}
\exp(-t\sigma_2\ln(2+2v))\exp(t\sigma_1\ln(2+2u)),
\end{equation}
whose magnitude is given by
\begin{equation}
|(\ref{eq23})|=\exp(-t\sigma_2\ln(2|1+v|))\exp(t\sigma_1\ln(2|1+u|)).
\end{equation}
a) For the term $(1+2v)/((u-v)(1+u+v))$, we can choose $t\gg 1$ such that
\begin{equation}
\textrm{dist}(\Gamma_0,-1-\Gamma_{-1})=\frac{\sqrt{\sigma_2}}{2}\geq \frac{\sqrt{L}}{4t^{1/3}}
\end{equation}
which is much better than in Proposition 4, where we had, see Figure~\ref{FigSteepPaths}
 \mbox{$\textrm{dist}(\Gamma_0,-1-\Gamma_{-1})=\delta t^{-1/3}$}.
Therefore the term $(1+2v)/((u-v)(1+u+v))$ does not create any problems.\\
b) Similarly, $\Gamma_{-1}$ can be chosen such that the maximum of $|1+u|$, for $u\in\Gamma_{-1}$, is obtained at $u=-1/2$, thus
\begin{equation}
e^{t\sigma_1\ln(2|1+u|)} \leq 1.
\end{equation}
c) $\Gamma_0$ can be chosen such that the minimum of $|1+v|$ for $v\in\Gamma_0\setminus\Gamma_{\rm vert}$, is obtained at $\Gamma_{\rm vert}$ for $\xi=\pm a$. A simple computation leads to
\begin{equation}\label{eq25}
e^{-t\sigma_2\ln(2|1+v|)} = e^{-t\sigma_2\ln(1+\sqrt{\sigma_2}+\Or(\sigma_2))}
= e^{-\tilde s_2^{3/2}(1+\Or(\e))/\sqrt{2}} \leq e^{-\tilde s_2^{3/2}/2}
\end{equation}
for $\e$ small enough.\\
d) Now we evaluate the integral over $\Gamma_{\rm vert}$. As considered in the $\xi$ variable, the prefactor $t^{-1/3}$ cancels out and $\xi$ varies over an interval of order one. Therefore, to estimate the integral it is enough to estimate the integrand. Since $\e$ is small, $\sigma_2$ is small too. Thus, $\Gamma_{\rm vert}$ is very close to $-1/2$ and we can apply Taylor expansion of the integrand. The term with the exponential in the $v$ variable becomes ($v=-1/2+\sqrt{\sigma_2}(1+\I \xi)/2$)
\begin{equation}\label{eq28}
\exp\big(t f_0(-1/2)+\tfrac16 t \sigma_2^{3/2} (1+\I \xi)^3+\tau_2 (t/2)^{2/3}\sigma_2(1+\I\xi)^2-\sigma_2^{3/2} t (1+\I \xi)+\Or(t\sigma_2^2)\big).
\end{equation}
By using $\sigma_2=\tilde s_2 t^{-2/3} 2^{-1/3}$ and computing the real values of the exponent, we get
\begin{equation}\label{eq29}
|(\ref{eq28})|\leq \exp\big(t f_0(-1/2)+ \tilde s_2^{3/2} 2^{-1/2}(-\tfrac56 -\tfrac12 \xi^2+\Or(\sqrt{\e}))+ \tfrac12 \tau_2 \tilde s_2 (1-\xi^2) \big).
\end{equation}
Here we have $s_2\geq L$, thus $\tilde s_2\geq s_2/2$ for large $L$ and $\tilde s_2^{3/2}\gg \tilde s_2$. Therefore, the integrand to be studied can be bounded by
\begin{equation}
e^{t f_0(-1/2)} e^{-\tilde s_2^{3/2}/2}
\end{equation}
for $L$ large enough and $\e$ small enough. The factor $e^{t f_0(-1/2)}$ is cancelled exactly with the one coming from the integrand in the $u$ variable.

For the case $\sigma_1=\sigma_2=0$, the analysis of Proposition~\ref{PropUniformConvergence} leads to the bound on the kernel $\widehat K_t^{\rm resc}$
\begin{equation}
(\ref{eq22})+\frac{1}{(2\pi)^2} \int_{\gamma_+}\dx w \int_{\gamma_-}\dx z\bigg|
\frac{e^{w^3/3+\tau_2 w^2}}{e^{z^3/3+\tau_1 z^2}} \frac{2w}{(z-w)(z+w)}\bigg|\leq c_6
\end{equation}
for some constant $c_6>0$, as soon as $t$ is large enough.

Putting together the results of a)-d), the kernel is bounded by $c_6$ times the factor $e^{-\tilde s_2^{3/2}/2}$. For $L$ large, $\tilde s_2\geq L/2$ and $\tilde s_2\geq s_2/\sqrt{2}$, therefore
\begin{equation}\label{eq31}
c_6 e^{-\tilde s_2^{3/2}/2} \leq c_6 e^{-\tfrac14\sqrt{L} s_2} \leq c_6 e^{-\tfrac18 \sqrt{L}(s_1+s_2)} \leq e^{-(s_1+s_2)}
\end{equation}
where we used $s_2\geq s_1$.

\textbf{Case $\sigma_1\geq \sigma_2$.} To obtain the bound for this case, we use a different expression for the kernel $\widehat{K}_t^{\rm resc}$, namely
\begin{eqnarray}\label{eq32}
\widehat{K}_t^{\rm resc}&=&(t/2)^{1/3}\frac{1}{(2\pi\I)^2}\oint_{\Gamma_0}\dx v \oint_{\Gamma_{-1}}\dx u \frac{1+2v}{(u-v)(1+u+v)}\\
&\times & \frac{\exp(t f_0(v)+(t/2)^{2/3}\tau_2 f_1(v)+(t/2)^{1/3}\tilde s_2 f_2(v))}
{\exp(t f_0(u)+(t/2)^{2/3}\tau_1 f_1(u)+(t/2)^{1/3}\tilde s_1 f_2(u)+f_3(u))}+I_2,\nonumber
\end{eqnarray}
where
\begin{eqnarray}\label{eqI2}
I_2&=&(t/2)^{1/3}\frac{-1}{2\pi\I}\oint_{\Gamma_0}\dx v e^{t(f_0(v)-f_0(-1-v))}e^{(t/2)^{2/3}(\tau_2 f_1(v)-\tau_1 f_1(-1-v))}\nonumber \\
&\times & e^{(t/2)^{1/3}(\tilde s_2 f_2(v)-\tilde s_1 f_2(-1-v))}e^{-f_3(-1-v)},
\end{eqnarray}
with the constraint $\Gamma_{-1}\subset -\Gamma_0$ instead of $-\Gamma_0\subset \Gamma_{-1}$. The term $I_2$ comes from the fact that, for any fixed $v$, the new constraint on the paths is obtained by deforming $\Gamma_{-1}$ and during this process one passes via a simple pole at $u=-1-v$, whose residue is $I_2$.

The analysis of the double integral term in (\ref{eq32}) is the same as in the previous case, where however $(u,s_1,\tau_1)$ play the role of $(v,s_2,\tau_2)$, so this time it is $\Gamma_{-1}$ which is modified instead of $\Gamma_0$ (symmetrically w.r.t.\ $-1/2$). We can then get as in (\ref{eq31}) the bound $\exp(-(s_1+s_2))/2$ and it remains to prove that $I_2$ is bounded by $\exp(-(s_1+s_2))/2$ too.

Denote $h_0(v)=f_0(v)-f_0(-1-v)$. It is given by $h_0(v)=-1+2 f_0(v)$. Therefore the regions where sign of $\Re(h_0(v)-h_0(-1/2))$ is positive and negative are again the ones of Figure~\ref{FigZeros}. In the case $\sigma_1=\sigma_2=0$, one can do essentially the asymptotic analysis made to obtain the estimate on the integral over $\Gamma_{-1}$ of Proposition~\ref{PropUniformConvergence} and we get that the integral is bounded in the $t\to\infty$ limit. The corrections to the limit expression are of just order $\Or(t^{-1/3},e^{-\mu t})$, for some $\mu>0$. But here we are in the case $s_1\in [L,\e t^{2/3}]$. The difference with respect to the case $\sigma_1=\sigma_2=0$ is a factor of magnitude
\begin{equation}
\exp(t \sigma_1 \ln(2|v|)-t\sigma_2 \ln(2|1+v|)),
\end{equation}
in the integrand. The $\Gamma_0$ used for the $\sigma_1=\sigma_2=0$ asymptotic analysis can be chosen such that, while going away from the critical point $v=-1/2$,\\
a) $|v|$ decreases, thus $\ln(2|v|)$ decreases,\\
b) $|1+v|$ increases, thus $-\ln(2|1+v|)$ decreases,\\
take for example $-1-\Gamma_{-1}$ of Figure~\ref{FigPathsModerate}.

Now we use the same trick as above, namely we modify $\Gamma_0$ only in the neighborhood of $v=-1/2$ as in Figure~\ref{FigPathsModerate} (just this time the distance to $v=-1/2$ is $\sqrt{\sigma_1}/2$ instead of $\sqrt{\sigma_2}/2$). We denote $\Gamma_{\rm vert}$ the vertical piece here too. Then, the contribution on $\Gamma_0\setminus \Gamma_{\rm vert}$ carries an extra term (as in (\ref{eq25}))
\begin{equation}\label{eq33}
e^{-t\sigma_1\ln(2|1+v|)}\leq e^{-\tilde s_1^{3/2}/2},
\end{equation}
for $\e$ small enough. Then, for $L$ large enough, $|(\ref{eq33})|\leq -e^{-c_7 \sqrt{L}(s_1+s_2)}$ for some $c_7>0$.

For the contribution of the integral over $\Gamma_{\rm vert}$, we set $v=-1/2+V$ and do Taylor expansion. Then set $V=\sqrt{\sigma_1}(1+\I\xi)/2$ with $\xi\in[-a,a]$, for some $a\in(1/\sqrt{3},\sqrt{3})$. The integral over $\Gamma_{\rm vert}$ is an integral over $[-a,a]$, which writes
\begin{eqnarray}\label{eq35}
& &(t/2)^{1/3} \frac{-1}{2\pi}\int_{-a}^{a}\dx \xi \sqrt{\sigma_1} e^{t\sigma_1^{3/2}
 (1+\I\xi)^3(1+\Or(\sqrt{\e}))/3}
e^{-(t/2)^{2/3}(\tau_1-\tau_2)\sigma_1(1+\I\xi)^2(1+\Or(\e))}\nonumber \\
&\times&e^{-(\sigma_1+\sigma_2)\sqrt{\sigma_1}(1+\I\xi)t(1+\Or(\sqrt{\e}))}
e^{\Or(\sqrt{\e})}.
\end{eqnarray}
We then use\\
a) $\Re((1+\I\xi)^3)=1-3\xi^2$,\\
b) $\Re((1+\I\xi)^2)=1-\xi^2$,\\
c) $\sqrt{\sigma_1}t^{1/3}\geq \sqrt{L}$,\\
d) $2/3\leq |1+\Or(\sqrt{\e})|\leq 2$, for $\e$ small enough,\\
to obtain that $|(\ref{eq35})|$ is bounded by
\begin{equation}\label{eq34}
\int_{-a}^{a}\dx \xi c_8 \sqrt{\tilde s_1} \exp\Big(t \sigma_1^{3/2}\big(\tfrac13-\xi^2-c_9(1-\xi^2)/\sqrt{L}\big)\Big) \exp\big(-\tfrac23(\sigma_1+\sigma_2)t\sqrt{\sigma_1}\big).
\end{equation}
The integral (\ref{eq34}) is bounded and, for $L$ large enough, the integrand is maximal at $\xi=0$. Thus
\begin{eqnarray}
(\ref{eq34}) &\leq& c_{10} \sqrt{\tilde s_1}\exp\big(\tfrac13 t\sigma_1^{3/2}-c_9/\sqrt{L}-\tfrac23 (\sigma_1+\sigma_2)\sqrt{\sigma_1} t\big) \nonumber \\
&\leq& \exp(-\tfrac16(\sigma_1+\sigma_2)\sqrt{\sigma_1} t)
\end{eqnarray}
for $L$ large enough. Reinserting the expressions for $\sigma_1$ and $\sigma_2$, we have
\begin{equation}
|(\ref{eq34})|\leq \exp\big(-c_{10}(\tilde s_1+\tilde s_2)t\sqrt{\tilde s_1}\big)
\leq \exp\big(-c_{11}(s_1+s_2)t\sqrt{L}\big)
\end{equation}
for $L$ large enough and some $c_{11}>0$. This bound is good enough to get $\exp(-(s_1+s_2))/2$ as bound for $L$ large enough, $\e$ small enough and $t$ large enough.
\end{proof}

\begin{prop}[Large deviations]\label{PropLargeDeviations}
Set $\e>0$, then for $t$ large enough we have
\begin{equation}
\left|\widehat{K}_t^{\rm resc}(\tau_1,s_1;\tau_2,s_2)\right|\leq e^{-(s_1+s_2)}
\end{equation}
for $(s_1,s_2)\in [-L,\infty)^2\setminus [-L,\e t^{2/3}]^2$.
\end{prop}
\begin{proof} One can do large deviations directly, but a shorter way is to use the result of the moderate deviations. As in the proof of Proposition~\ref{PropModerateDeviations} we use the notation, $\sigma_i=\tilde s_i t^{-2/3} 2^{-1/3}$, $i=1,2$.

\textbf{Case $\sigma_1\leq \sigma_2$.} The term linear in $t$ in the exponential is \mbox{$\exp(t f_{0,\sigma_2}(v)-t f_{0,\sigma_1}(u))$}, where $f_{0,\sigma}(v)=f_0(v)-\sigma\ln(2+2v)$. To obtain the bound we just remark that
\begin{equation}
f_{0,\sigma_2}(v)=f_{0,\e/2}(v)-(\sigma_2-\e/2)\ln(2+2v).
\end{equation}
We take $\Gamma_0$ to be the one used for moderate deviations with $\sigma_2=\e/2$. $\Gamma_0$ satisfies $|1+v|\geq 1/2+\sqrt{\e/2}/2$. $\sigma_2\geq \e$ implies $\sigma_2-\e/2\geq \sigma_2/2$. Therefore, for $\e$ small enough and $t$ large enough,
\begin{equation}\label{eq4.38}
|\exp(-t(\sigma_2-\e/2)\ln(2+2v))| \leq \exp(-\tfrac12 t \sigma_2\ln(1+\sqrt{\e/2})) \leq \exp(-c_{12} t^{1/3} s_2).
\end{equation}
The integral (\ref{eq6}) with $f_0(v)=f_{0,\e/2}(v)$ is finite by the
 same argument as for the moderate deviations. The extra factor
 (\ref{eq4.38}) together with \mbox{$\tilde s_2\geq (\tilde s_1+\tilde
 s_2)/2$} leads to the bound $\exp(-(s_1+s_2))$ for $t$ large enough.

\textbf{Case $\sigma_1\geq \sigma_2$.} Using the representation as in the moderate deviation case, we have, with respect to $\sigma_1=\e/2$, the extra factor
\begin{equation}
\exp(\tfrac12t\sigma_1\ln(1-\sqrt{\e/2})) \leq \exp(-c_{13} t^{1/3} s_1),
\end{equation}
from which we get the bound $\exp(-(s_1+s_2))$ as before.
\end{proof}

\begin{prop}[Uniform convergence on bounded sets]\label{PropUniformConvergenceBinomial}
Fix any $L>0$ and $x_i,s_i$ with the above rescaling. Then, uniformly for $(s_1,s_2)\in [-L,L]^2$,
\begin{eqnarray}
& &\lim_{t\to\infty} (t/2)^{1/3} 2^{x_2-x_1}\binom{x_1-x_2-1}{n_2-n_1-1} \nonumber \\
&=&\frac{1}{\sqrt{4\pi (\tau_2-\tau_1)}}\exp\left(-\frac{(\tilde s_2-\tilde s_1)^2}{4 (\tau_2-\tau_1)}\right) \Id(\tau_2>\tau_1).
\end{eqnarray}
\end{prop}
\begin{proof}
It is a special case of the first part of Proposition 5.1 of~\cite{BFP06}, where $p$ is chosen such that $\kappa=2^{-1/3}$ and $(r_i,s_i)$ are replaced by $(\tau_i,\tilde s_i)$.
\end{proof}

\begin{prop}\label{PropBoundBinomial}
For any $s_1,s_2\in \R$ and $\tau_2-\tau_1>0$ fixed, the bound
\begin{eqnarray}
(t/2)^{1/3} 2^{x_2-x_1}\binom{x_1-x_2-1}{n_2-n_1-1}\leq c_{12} e^{-|\tilde s_2-\tilde s_1|}
\end{eqnarray}
holds for $t$ large enough and $c_{12}$ independent of $t$.
\end{prop}
\begin{proof}
It is a special case of the first part of Proposition 5.5 of~\cite{BFP06}, where $p$ is chosen such that $\kappa=2^{-1/3}$ and $(r_i,s_i)$ are replaced by $(\tau_i,\tilde s_i)$.
\end{proof}

\section{About the other three regions}\label{SectThreeRegions}
\subsubsection*{(1) Constant density region.}
To obtain the result in the constant density region we consider the scaling
\begin{eqnarray}
n_i&=&[\alpha t/4+\tau_i (t/2)^{2/3}],\nonumber \\
x_i&=&[(1-\alpha)t/2-2\tau_i (t/2)^{2/3}-s_i(t/2)^{1/3}]
\end{eqnarray}
with $\alpha>1$ fixed. The rescaled and conjugate kernel is as before
\begin{equation}
K_t^{\rm resc}(\tau_1,s_1;\tau_2,s_2)=K_t(n_1,x_1;n_2,x_2) (t/2)^{1/3} 2^{x_2-x_1}.
\end{equation}
The binomial term is easily estimated and controlled. The main term $\widehat K_t^{\rm resc}$ is given by the formula (\ref{eq6}), with $\tilde s_i = s_i$, $f_1,f_2,f_3$ as in (\ref{eq7}), and the new $f_0$ is
\begin{equation}
f_0(v)=-v+\frac{2-\alpha}{4}\ln(1+v)-\frac{\alpha}{4} \ln(-v).
\end{equation}
The two critical points $v_-,v_+$ of $f_0$ are now distinct, namely \mbox{$v_-=-\alpha/2 < -1/2=v_+$}. The constraint between the integration paths $-1-\Gamma_0 \subset \Gamma_{-1}$ can not be satisfied if we want to choose $\Gamma_0$ and $\Gamma_{-1}$ optimally, i.e., passing by $v_+$ and $v_-$ respectively. For the analysis, one considers another representation of $\widehat K_t^{\rm resc}$, the same used in (\ref{eq32}). The first term is as before but with the constraint $\Gamma_{-1}\subset -1-\Gamma_0$ and the second is the residue at $u=-1-v$, namely equal to $I_2$ in (\ref{eqI2}).

The first term is now controlled by choosing optimal paths for $f_0(v)$ and $-f_0(u)$, which pass by $v_+$ and $v_-$ respectively. $f_0(v_+)<f_0(v_-)$, thus the first term is of order $\Or(\exp(t f_0(v_+)-t f_0(v_-)))=\Or(e^{-a t})$ for some $a>0$. In particular, for $\alpha>2$ the first term vanishes identically (for $t$ large enough), and as $\alpha\searrow 1$, the first term is $\Or(e^{-t(\alpha-1)^3/12})$.

The second term is just $I_2$, up to some $2^{1/3}$ factors due to the slightly different rescaling, the same kernel appearing in (5.5) of~\cite{BFPS06}, where we already proved the pointwise convergence. The moderate and large deviations are the ones of $I_2$ in (\ref{eqI2}) analyzed in Propositions~\ref{PropModerateDeviations} and~\ref{PropLargeDeviations}. In the $t\to\infty$ limit one then obtains
\begin{equation}
\lim_{t\to\infty}K_t^{\rm resc}(\tau_1,s_1;\tau_2,s_2) = 2^{-1/3}K_{{\cal A}_1}(2^{-2/3}\tau_1,2^{-1/3}s_1;2^{-2/3}\tau_2,2^{-1/3}s_2)
\end{equation}
with $K_{{\cal A}_1}$ is the kernel of the Airy$_1$ process.

\subsubsection*{(2) Linearly decreasing density region.}
To obtain the result in the linearly decreasing density region we consider the scaling
\begin{eqnarray}
n_i&=&[\alpha t/4+\tau_i (t/2)^{2/3}],\nonumber \\
x_i&=&\left[(1-\sqrt{\alpha})t-\frac{2\tau_i}{\sqrt{\alpha}} (t/2)^{2/3}+\frac{\tau_i^2}{\alpha^{3/2}}(t/2)^{1/3}-s_i(t/2)^{1/3}\right]
\end{eqnarray}
with $0<\alpha<1$ fixed. The rescaled and conjugate kernel is
\begin{equation}
K_t^{\rm resc}(\tau_1,s_1;\tau_2,s_2)=K_t(n_1,x_1;n_2,x_2) (t/2)^{1/3} \frac{(\sqrt{\alpha}/2)^{n_2-n_1}}{(1-\sqrt{\alpha}/2)^{x_2+n_2-x_1-n_1}}.
\end{equation}
The main term of the kernel $\widehat K_t^{\rm resc}$ writes as (\ref{eq6}) with $\tilde s_i=s_i-\tau_i^2/\alpha^{3/2}$, $f_3$ as in (\ref{eq7}), and
\begin{eqnarray}
f_0(v)&=&-v+(1-\sqrt{\alpha}+\alpha/4)\ln(1+v)-(\alpha/4)\ln(-v),\nonumber \\
f_1(v)&=&(1-2/\sqrt{\alpha})\ln(1+v)-\ln(-v)+\ln(\sqrt{\alpha}/2)\nonumber \\
& &-(1-2/\sqrt{\alpha})\ln(1-\sqrt{\alpha}/2),\nonumber\\
f_2(v)&=&-\ln(1+v)+\ln(1-\sqrt{\alpha}/2).
\end{eqnarray}
The function $f_0(v)$ has a double critical point at $v=-\sqrt{\alpha}/2$. The factor $1+u+v=1-\sqrt{\alpha}$ at the critical point and the paths $\Gamma_0$ and $\Gamma_{-1}$ can be chosen such that $1+u+v$ remains uniformly bounded away from $0$. The leading term of the integral comes from the neighborhood of the critical point. There, one applies the following change of variables,
\begin{equation}
v=-\frac{\sqrt{\alpha}}{2}+\frac{\alpha^{1/6}(2-\sqrt{\alpha})^{1/3}}{2^{2/3}t^{1/3}}V,\quad
u=-\frac{\sqrt{\alpha}}{2}+\frac{\alpha^{1/6}(2-\sqrt{\alpha})^{1/3}}{2^{2/3}t^{1/3}}U.
\end{equation}
Set $S_h=\alpha^{-2/3}(2-\sqrt{\alpha})^{-1/3}$ and $S_v=\alpha^{1/6}(2-\sqrt{\alpha})^{-2/3}$. Then, the leading term in the main term of the kernel becomes
\begin{equation}
\widehat K_t^{\rm resc}(\tau_1,s_1;\tau_2,s_2)\simeq \frac{S_v}{(2\pi\I)^2}\int\dx V\int \dx U \frac{1}{U-V} \frac{e^{V^3/3+\tau_2 S_h V^2-\tilde s_2 S_v V}}{e^{U^3/3+\tau_1 S_h U^2-\tilde s_1 S_v U}}.
\end{equation}
Thus
\begin{equation}
K_t^{\rm resc}(\tau_1,s_1;\tau_2,s_2)\to S_v K_{{\cal A}_2}(S_h\tau_1,S_v s_1;S_h\tau_2,S_v s_2)
\end{equation}
as $t\to\infty$. By adequate control for moderate and large deviations, one proves (\ref{eqCurved}).

\subsubsection*{(3) Finite distance from the right-most particle.}
From the discussion on the initial condition, in particular from (\ref{eq2.12}), it follows that the asymptotic result is unchanged if one considers step initial conditions instead of our initial conditions. In~\cite{JN06} the case of step initial conditions was analyzed in a closely related model (a kind of discrete time TASEP but from the growth point of view). For step initial conditions, we have
\begin{equation}
K_t(n_1,x_1;n_2,x_2)=-\binom{x_1-x_2-1}{n_2-n_1-1}\Id_{[n_2>n_1]}+\widehat K_t(n_1,x_1;n_2,x_2)
\end{equation}
with
\begin{equation}
\widehat K_t(n_1,x_1;n_2,x_2)=\sum_{k=0}^{n_2-1}\Psi^{n_1}_{n_1-n_2+k}(x_1) \Phi^{n_2}_k(x_2),
\end{equation}
where
\begin{equation}
\Psi^n_k(x)=\frac{e^{-t}t^{x+2n}}{(x+2n)!} C_k(x+2n,t),\quad \Phi^n_l(y)=C_l(y+2n,t),
\end{equation}
the $C_k$ being the Charlier orthogonal polynomials~\cite{KS96}. This is obtained in the same way as in Appendix B of~\cite{BFPS06}. $\Psi^n_k(z)$ is the same as (B.7) of~\cite{BFPS06} with $z-k$ replaced by $z=x+2n$, and consequently the matrix $S_{k,l}$ becomes the identity matrix.

The Charlier polynomials converge to the Hermite polynomials $H_k$ as follows
\begin{equation}\label{eq5.14}
\lim_{t\to\infty} (2t)^{k/2} C_k(t-\sqrt{2t}\sigma,t)=(-1)^k H_k(-\sigma)=H_k(\sigma).
\end{equation}
The scaling we have to use is
\begin{equation}
n_i,\quad x_i=[t-\sqrt{2t}s_i]
\end{equation}
and the kernel rescaled as
\begin{equation}
K_t^{\rm resc}(n_1,s_1;n_2,s_2)=\sqrt{2t}\frac{e^{-s_2^2/2+s_1^2/2}}{t^{n_2/2-n_1/2}} K_t(n_1,x_1;n_2,x_2).
\end{equation}

It is easy to see that the binomial contribution converges to
\begin{equation}\label{eq5.17}
-\frac{e^{(s_1^2-s_2^2)/2}2^{(n_2-n_1)/2}}{(n_2-n_1-1)!}(s_2-s_1)^{n_2-n_1-1}\Id_{[s_2>s_1]}.
\end{equation}
Also, by (\ref{eq5.14}), we have
\begin{equation}
\lim_{t\to\infty} \Psi^n_k(x_i)=\frac{e^{-s_i^2}}{(2t)^{k/2}\sqrt{2\pi t}}H_k(s_i),\quad
\lim_{t\to\infty} \Phi^n_k(x_i)=\frac{(t/2)^{k/2}}{k!} H_k(s_i).
\end{equation}
The kernel is a finite sum, thus
\begin{equation}\label{eq5.18}
\lim_{t\to\infty}\widehat K_t^{\rm resc}(n_1,s_1;n_2,s_2)= e^{-(s_1^2+s_2^2)/2} \sum_{j=-n_2}^{-1} \sqrt{\frac{(n_1+j)!}{(n_2+j)!}}h_{n_2+j}(s_1) h_{n_1+j}(s_2)
\end{equation}
where $h_k(s)=\pi^{-1/4} k!^{-1/2} 2^{-k/2} H_k(s)$.

(\ref{eq5.17}) plus (\ref{eq5.18}) gives
\begin{equation}
\lim_{t\to\infty} K_t^{\rm resc}(n_1,s_1;n_2,s_2)=K^{\rm GUE}(n_2,s_2;n_1,s_1)
\end{equation}
with $K^{\rm GUE}$ the kernel defined in Definition 1.2 of~\cite{JN06}. (Here we just order the entries differently).

\appendix

\section{Explicit form of the limit kernel}\label{AppExplicitKernel}

\subsubsection*{Transition kernel in terms of Airy functions}
Let us denote
\begin{equation}
\tilde s_i=s_i-\min\{0,\tau_i\}^2,\quad \hat s_i=s_i+\max\{0,\tau_i\}^2.
\end{equation}
Then
\begin{equation}\label{eqA2}
K_\infty(\tau_1,s_1;\tau_2,s_2) = K_0(\tau_1,s_1;\tau_2,s_2)+K_1(\tau_1,s_1;\tau_2,s_2)+K_2(\tau_1,s_1;\tau_2,s_2)
\end{equation}
where
\begin{equation}\label{eqA3}
K_0(\tau_1,s_1;\tau_2,s_2)= -\frac{e^{\tfrac23\tau_2^3+\tau_2\tilde s_2}}{e^{\tfrac23\tau_1^3+\tau_1 \tilde s_1}}
\frac{1}{\sqrt{4\pi(\tau_2-\tau_1)}}\exp\left(-\frac{(\tilde s_2-\tilde s_1)^2}{4(\tau_2-\tau_1)}\right) \Id_{[\tau_2>\tau_1]},
\end{equation}
\begin{equation}\label{eqA4}
K_2(\tau_1,s_1;\tau_2,s_2) =
\int_0^\infty\dx \lambda e^{\lambda(\tau_2-\tau_1)} \Ai(\hat s_2+\lambda)\Ai(\hat s_1+\lambda),
\end{equation}
and
\begin{equation}\label{eqA5}
K_1(\tau_1,s_1;\tau_2,s_2)=
\int_0^\infty\dx \lambda e^{\lambda(\tau_2+\tau_1)} \Ai(\hat s_2+\lambda)\Ai(\hat s_1-\lambda).
\end{equation}
Equivalently, one can see that
\begin{eqnarray}\label{eqA6}
& &\hspace{-2.5em}K_1(\tau_1,s_1;\tau_2,s_2)=-\int_{-\infty}^0\dx \lambda e^{\lambda(\tau_2+\tau_1)} \Ai(\hat s_2 +\lambda)\Ai(\hat s_1-\lambda)\nonumber \\
& &+2^{-1/3}\Ai\left(2^{-1/3}(\tilde s_1+\tilde s_2+\tfrac12 (\tau_1-\tau_2)^2)\right) e^{-\tfrac12(\tau_1+\tau_2)(\hat s_2-\hat s_1)}
\end{eqnarray}

\section{Trace class and transition kernel}\label{AppTraceClass}
\begin{prop}\label{PropTraceClass}
The Fredholm determinant
\begin{equation}
\det(\Id-\chi_s K_\infty \chi_s)_{\Hilb},
\end{equation}
with $K_\infty$ given in (\ref{eqKCompleteInfinity}) or (\ref{eqA2})-(\ref{eqA5}), is well defined, because there exists a conjugate kernel of $\chi_s K_\infty \chi_s$ which is trace-class on $\Hilb=L^2(\{\tau_1,\ldots,\tau_m\}\times \R)$.
\end{prop}
\begin{proof}
In this proof, let us choose a $T_0$ such that $-T_0 < \tau_1 <\tau_2 < \ldots < \tau_m < T_0$. Denote by $K^{\rm conj}$ a conjugate of $\chi_s K_\infty \chi_s$. Let $P_k$ be the projector onto the space $\{f \in \Hilb | f(\tau_l,x)=0\textrm{ for }l\neq k\}$ and $K^{\rm conj}_{k,l}=P_k K^{\rm conj} P_l$. Then
\begin{equation}
\|K^{\rm conj}\|_1 \leq \sum_{k,l=1}^m \|K^{\rm conj}_{k,l}\|_1.
\end{equation}

From (\ref{eqA2}) we have
\begin{equation}
K^{\rm conj}_{k,l}(\tau_k,x;\tau_l,y)=\Id_{[x\geq s_k]}\Id_{[y\geq s_l]} \frac{\rho(\tau_k,x)}{\rho(\tau_l,y)} \sum_{n=0,1,2} K_n(\tau_k,x;\tau_l,y)
\end{equation}
where the conjugation function $\rho(\tau,x)\neq 0$ will be specified later.

The formula defining $K_0(\tau_k,x;\tau_l,y)$ is particularly nice in
 the \mbox{$\tilde s_k=s_k-[\tau_k]_-^2$} variables (with $[x]_-=x$ for
 $x\leq 0$ and $0$ otherwise). Thus we use the variables $\tilde s_k$
 instead of $s_k$. This is just a shift of the coordinate at the
 corresponding ``time'' $\tau_k$. Thus, if we prove that $\widetilde
 K^{\rm conj}_{k,l}(\tau_k,x;\tau_l,y)
=K^{\rm conj}_{k,l}(\tau_k,x+[\tau_k]_-^2;\tau_l,y+[\tau_l]_-^2)$
is trace-class for all $\tilde s_k$, $k=1,\ldots,m$, bounded from below, then $K^{\rm conj}_{k,l}$ will also be trace-class for all $s_k$ bounded from below.

Therefore, we now work with the $\widetilde K^{\rm conj}$ kernels and choose the conjugation functions ($\tilde \rho(\tau_k,x)=\rho(\tau_k,x+[\tau_k]_-^2)$) to be
\begin{equation}
\tilde \rho(\tau_k,x)=(1+x^2)^{2k} e^{\tau_k x + \tfrac23 \tau_k^3}.
\end{equation}
We analyze separately the three parts of the kernel. Let $\widetilde K_n(\tau_k,x;\tau_l,y)=K_n(\tau_k,x+[\tau_k]_-^2;\tau_l,y+[\tau_l]_-^2)$, $n=0,1,2$.\\[0.5em]
\noindent {\bf Part a) $\widetilde K_0(\tau_k,x;\tau_l,y)$.} We have
\begin{eqnarray}
& &\widetilde K_0(\tau_k,x;\tau_l,y) \frac{\tilde \rho(\tau_k,x)}{\tilde \rho(\tau_l,y)}\Id_{[x\geq \tilde s_k]} \Id_{[y\geq \tilde s_l]}\\
&=& -\frac{\Id_{[\tau_l>\tau_k]}}{\sqrt{4\pi(\tau_l-\tau_k)}} \Id_{[x\geq \tilde s_k]}\Id_{[y\geq \tilde s_l]} \exp\left(-\frac{(y-x)^2}{4(\tau_l-\tau_k)}\right)\frac{(1+x^2)^{2k}}{(1+y^2)^{2l}}.
\nonumber
\end{eqnarray}
In Lemma A.2 of~\cite{BFP06}, we proved that the operator with above kernel is trace-class on $L^2(\R)$. (Recall that $\tau_l>\tau_k$ if and only if $l>k$).\\[0.5em]
\noindent {\bf Part b) $\widetilde K_2(\tau_k,x;\tau_l,y)$.} We have
\begin{equation}
\widetilde K_2(\tau_k,x;\tau_l,y) \frac{\tilde \rho(\tau_k,x)}{\tilde \rho(\tau_l,y)}\Id_{[x\geq \tilde s_k]} \Id_{[y\geq \tilde s_l]} = \int_{\R}\dx\lambda A_1(x,\lambda) A_2(\lambda,y)
\end{equation}
with
\begin{equation}
A_1(x,\lambda)=\Id_{[x\geq \tilde s_k]}\Id_{[\lambda\geq 0]} \tilde \rho(\tau_k,x)e^{-\tau_k \lambda} \Ai(x+\lambda+\tau_k^2)
\end{equation}
and
\begin{equation}
A_2(\lambda,y)=\Id_{[\lambda\geq 0]}\Id_{[y\geq \tilde s_l]} \frac{e^{\tau_l \lambda}}{\tilde \rho(\tau_l,y)}
\Ai(y+\lambda+\tau_l^2).
\end{equation}

Then we use $\|A_1 A_2\|_1 \leq \|A_1\|_2\, \|A_2\|_2$. Thus we have just to prove that $A_1$ and $A_2$ are Hilbert-Schmidt operators. This is easy to see, since
\begin{eqnarray}
\|A_1\|_2^2 &=& \int_{\R^2}\dx x\dx \lambda |A_1(x,\lambda)|^2 \\
&=& \int_{\tilde s_k}^\infty\dx x \int_0^\infty \dx\lambda \tilde \rho(\tau_k,x)^2 e^{-2\tau_k \lambda} |\Ai(x+\lambda+\tau_k^2)|^2\nonumber \\
&\leq& C(T_0,\tilde s_k)<\infty \nonumber
\end{eqnarray}
because the integrand is bounded, and for large $x$ and $\lambda$ the decay is super-exponential due to the Airy function ($\Ai(z)\simeq e^{-\tfrac23 z^{3/2}}$ for $z\gg 1$). Similarly one shows that $\|A_2\|_2<\infty$.\\[0.5em]
\noindent {\bf Part c) $\widetilde K_1(\tau_k,x,\tau_l,y)$.} We have
\begin{equation}
\widetilde K_1(\tau_k,x,\tau_l,y) \frac{\tilde \rho(\tau_k,x)}{\tilde \rho(\tau_l,y)}\Id_{[x\geq \tilde s_k]} \Id_{[y\geq \tilde s_l]} = \int_{\R}\dx\lambda B_1(x,\lambda) B_2(\lambda,y)
\end{equation}
with
\begin{equation}
B_1(x,\lambda)=\Id_{[x\geq \tilde s_k]}\Id_{[\lambda\geq 0]} e^{\tfrac23 \tau_k^3}(1+x^2)^{2k} e^{\tau_k x} e^{3 T_0\lambda} \Ai(x+\lambda+\tau_k^2)
\end{equation}
and
\begin{equation}
B_2(\lambda,y)=\Id_{[\lambda\geq 0]}\Id_{[y\geq \tilde s_l]} e^{-\tfrac23 \tau_l^3} \frac{1}{(1+y^2)^{2l}} f(\lambda) g(\lambda,y)
\end{equation}
with $f(\lambda)=e^{(\tau_l+\tau_k-2T_0)\lambda}$ and $g(\lambda,y)=e^{-\tau_l y}e^{-T_0\lambda} \Ai(y-\lambda+\tau_l^2)$.

We need some estimates now. Since $\tau_l+\tau_k-2T_0<2\tau_m-2T_0$, we have
\begin{equation}\label{eqA10}
|f(\lambda)|\leq e^{-\mu \lambda}
\end{equation}
for $\mu=2(T_0-\tau_m)>0$. Moreover,
\begin{equation}
|g(\lambda,y)| = e^{-T_0\lambda} e^{-\tau_l y} | \Ai(y+\tau_l^2-\lambda)|.
\end{equation}
Setting $z=y+\tau_l^2$ and $c_1=e^{\tau_l^3}$, we get
\begin{equation}
|g(\lambda,y)| \leq c_1 e^{-\tau_l z} e^{-T_0 \lambda} |\Ai(z-\lambda)|.
\end{equation}
The first case is $z\leq \lambda$. There, $|\Ai(z-\lambda)|\leq 1$, thus
\begin{equation}\label{eqA11a}
|g(\lambda,y)| \leq c_2.
\end{equation}
The second case is $z \geq \lambda$ (recall that $\lambda\geq 0$). There
\begin{equation}\label{eqA11b}
|g(\lambda,y)| \leq c_1 e^{T_0(z-\lambda)}\Ai(z-\lambda) \leq c_3
\end{equation}
because $\max_{x\geq 0} e^{T_0 x} \Ai(x) = c_3<\infty$ due to the super-exponential decay of $\Ai(x)$ for large $x$. Thus by (\ref{eqA11a}) and (\ref{eqA11b}) we conclude that, for all $\lambda\geq 0$ and $y\geq \tilde s_l$, there exists a constant $c_4$ such that $|g(\lambda,y)|\leq c_4$.

The inequality $\|B_1\|_2<\infty$ is similar to the $\|A_1\|_2$ case (use the decay of the Airy function).
To see that $\|B_2\|_2<\infty$, we use the bound (\ref{eqA10}) to control the behavior in $\lambda$, $|g|$ is just  bounded by a constant and the decay in $y$ is controlled by the $(1+y^2)^{-2l}$ term.

In parts a), b) and c) we proved that all the kernel elements are trace-class on $L^2(\R)$ and this ends the proof of Proposition~\ref{PropTraceClass}.
\end{proof}


\end{document}